\documentclass[fleqn,usenatbib]{mnras}

\usepackage{newtxtext,newtxmath}

\usepackage[T1]{fontenc}
\usepackage{ae,aecompl}

\usepackage{graphicx}	
\usepackage{amsmath}	
\usepackage{amssymb}	
\hypersetup{draft}

\newcommand{\gaia}{\emph{Gaia}}

\makeatletter
\newcommand{\@todonotes@enable}{0}
\newcommand{\@todonotes@inline}{1}
\makeatother
\input{notes.tex}

\definecolor{lightgreen}{rgb}{0.0, 1., 0.0}

\title[\gaia\ anisotropy profiles of globular clusters]{The orbital anisotropy profiles of nearby globular clusters from \gaia\ Data Release 2}

\author[Jindal et al.]{
Abhinav Jindal\thanks{E-mail: abhinav.jindal@mail.utoronto.ca},
Jeremy J. Webb, 
and Jo Bovy
\\
Department of Astronomy and Astrophysics, University of Toronto, 50 St. George Street, Toronto, ON, M 5S 3H4, Canada\\
}

\date{Accepted XXX. Received YYY; in original form ZZZ}

\pubyear{2019}

\begin{document}
\label{firstpage}
\pagerange{\pageref{firstpage}--\pageref{lastpage}}
\maketitle

\begin{abstract}
 \gaia\ Data Release 2 provides a wealth of data to study the internal structure of nearby globular clusters. We use this data to investigate the internal kinematics of 11 nearby globular clusters, with a particular focus on their poorly-studied outer regions. We apply a strict set of selection criteria to remove contaminating sources and create pure cluster-member samples over a significant fraction of the radial range of each cluster. We confirm previous measurements of rotation (or a lack thereof) in the inner regions of several clusters, while extending the detection of rotation well beyond where it was previously measured and finding a steady decrease in rotation with radius. We also determine the orbital anisotropy profile and determine that clusters have isotropic cores, are radially anisotropic out to $\approx$ 4 half-light radii or $35\%$ of their limiting radii, and are then isotropic out to the limits of our datasets. We detect for the first time the presence of radial anisotropy in M 22, while confirming previous detections of radial anisotropy in 47 Tuc, M 3, M 13, M 15, and $\omega$ Cen's innermost regions. The implications of these measurements are that clusters can be separated into two categories: 1) clusters with observed radial anisotropy that likely formed tidally under-filling or are dynamically young, and 2) clusters that are primarily isotropic that likely formed tidally filling or are dynamically old.
\end{abstract}

\begin{keywords}
astrometry --- globular clusters: general --- methods: data analysis --- stars: kinematics and dynamics
\end{keywords}



\section{Introduction}
\label{sec:intro}

Globular clusters (GCs) are spherical collections of old, metal poor stars with density and velocity dispersion increasing greatly toward their centres. The literature is rich with observational studies, several dating back many decades, of the photometric, spectroscopic, and kinematic properties of GCs \citep[e.g][]{hartwick76, gunn79, mcclure86, peterson89,chaboyer95, mcLaughlin05}. However, all of these studies were restricted by the fact that proper motion data was unavailable for GC stars, making it nearly impossible to confirm cluster membership. Proper motions required nearly a century of baseline time for detection \citep{rees93}, and even then, only an aggregate proper motion of the entire cluster could be characterized. Hence, models of how GCs form and evolve were difficult to constrain.

It has not been until recently that telescopes and detectors have advanced enough such that the internal kinematic properties of GCs can be studied with precision. Using line-of-sight velocity data collected from spectroscopy, numerous studies have detected significant rotation in many GCs \citep[e.g.][]{lane11, kimmig15, boberg17, kacharov14}. These detections were surprising, as the surface density profiles of most GCs could be represented as an isotropic, non-rotating collection of stars \citep{king62, king66, wilson75}. Rotation imparted onto a newly formed GC due to the collapse of the giant molecular cloud from which it formed was thought to only affect the cluster's early evolution, possibly being a contributing factor to both early cluster mass loss and setting its ellipticity \citep{lagoute96,longaretti96}. Over time, cluster rotation would dissipate due to the loss of angular momentum through internal relaxation processes \citep[e.g.][]{tiongco17}. 

Requiring a shorter baseline than past telescopes because of its higher angular resolution, the Hubble Space Telescope (HST) opened the door for investigating the internal kinematics of GCs using the proper motions of member stars. For example, the addition of proper motion measurements revealed significant rotation in $\omega$ Cen \citep{bellini18}. Furthermore, radial velocities and proper motions have allowed for the detection of a slightly radially-biased velocity dispersion in several GCs \citep{richer13,bellini15,watkins15}. However, studies with HST are limited to the inner few arcmin of the clusters \citep{watkins15}, which typically do not probe distances beyond the half-light radius. Being able to measure the complete anisotropy profile of a cluster at larger radii is extremely informative, as it has been shown to be connected to a cluster's properties at formation, dynamical history, and the tidal field of its host galaxy \citep{tiongco16}. Assuming clusters form compact and with stars having an isotropic distribution of orbits, a non-negligible degree of radial anisotropy in stellar velocities is expected to develop in the outer regions from the cluster expanding due to stellar evolution and the scattering of stars from the core to highly eccentric orbits \citep{zocchi16}. Initially extended clusters that are tidally filling are not expected to develop strong radial anisotropy as stars that reach radial orbits will quickly escape the cluster \citep{tiongco16}. For dynamically old clusters, relaxation and mass loss via tidal stripping leave the remaining population of stars to be primarily isotropic \citep{zocchi16,tiongco16}. Knowing the orbital anisotropy profile is also important for the dynamical modeling of globular clusters and constraining their total masses, including possible contributions from dark matter \citep{zocchi17, henaultbrunet2019, claydon19}. For example, tangential anisotropy is expected to exist in clusters that are strongly rotating \citep{vesperini2014} or have been recently subjected to a tidal shock via dark matter substructure \citep{webb2019}.

The recent second data release of \gaia\ (DR2) has drastically improved the available kinematic data on globular clusters. The extensive and accurate astrometric database in \gaia\ DR2 contains proper motions for tens of thousands of sources within GCs and questions regarding their internal kinematics and how they relate to their formation and dynamical history are now worth revisiting. N-body simulations of GCs have shown that a complete understanding of the internal dynamics of GCs can provide strong constraints on GC formation scenarios \citep{mastrobuono13, vesperini13, henault15, mastrobuono16}.

\citet{milone18} focused on studying the kinematics of one GC in particular: 47 Tuc. They present radial profiles of the tangential proper motion and velocity dispersion out to 18 arcmin from the cluster centre (6 half-light radii $r_h$), considering only red giant branch sources with good astrometry. They do not attempt to filter out contaminating sources, though this is not expected to significantly affect the results for this cluster in its inner parts where its star counts dominate over those of contaminants. The authors confirm that 47 Tuc rotates on the plane of the sky and has an anisotropy profile that is line with theoretical expectations. More specifically, the core of 47 Tuc is isotropic, with the degree of radial anisotropy increasing between $\sim\!\!1$ to 3 $r_h$ and then decreasing towards isotropy again from $\sim\!\!3$ to 5 $r_h$, likely due to the presence of external tides. 

\citet{bianchini18} searched for rotation signatures in the inner regions of 51 GCs and tabulated the tangential component of the proper motions as a function of radius. They attempt to filter out contaminating sources by considering the parallax, the proximity of the source to the cluster's isochrone on a colour-magnitude diagram (CMD), and the average proper motion. The authors detect rotation in 22 clusters with greater than 2$\sigma$ confidence. \citet{bianchini18} also find evidence for a correlation between the degree of rotation and the relaxation time GCs, which indicates the importance of angular momentum in cluster formation.

\citet{baumgardt18b} generated a comprehensive catalog of mean proper motions for nearly all of the GCs in the Milky Way to complement the analysis presented in \citet{gaiahelmi18}, and they tabulated the internal velocity dispersion profiles. By cross-correlating \gaia\ objects with objects whose line-of-sight velocities are known from ESO and Keck spectra, they are able to use line-of-sight velocities as a way to filter out contaminating objects. The authors apply additional filters by considering parallax, proximity to the stellar isochrone on a CMD, proper motion, and proper motion errors. 

Each of these studies using \gaia\ DR2 have primarily focused on the innermost regions of the GCs and they do not probe out to distances close to the tidal radii of the clusters, because contamination from foreground and background stars---which we will refer to simply as the ``background''---increases strongly in the sparsely-populated outer regions. More sophisticated filtering is necessary to remove the background and study the outer regions of GCs. Only the recent work by \citet{deBoer19} has used \gaia\ data to identify cluster stars in the outer regions to determine the density profiles of 81 GCs out to each cluster's limiting radius $r_L$. Since the density profile of the background is quite smooth compared to that of the cluster, filtering out the background is not that important for the density measurement.

The kinematics of stars at large clustercentric distances are important for understanding the interaction of cluster dynamics with the surrounding tidal field, which is responsible for stellar mass loss and limiting cluster expansion. When studying the outer regions of GCs with \gaia\ data, the cluster's stellar density diminishes and contamination from background sources becomes significant. In this investigation, we determine the internal kinematics in the poorly-studied outer regions of 11 nearby GCs. We combine several types of filtering methods to thoroughly eliminate background contamination, which is crucial at such large clustercentric radii. In Sec.~\ref{sec:selection}, we detail our selection criteria. We present the kinematic profiles of each GC in Sec.~\ref{sec:resultsdiscussion} and make concluding remarks in Sec.~\ref{sec:conclusion}.

\section{Data} \label{sec:selection}
\subsection{Member selection}

\begin{figure}
\includegraphics[width=\columnwidth]{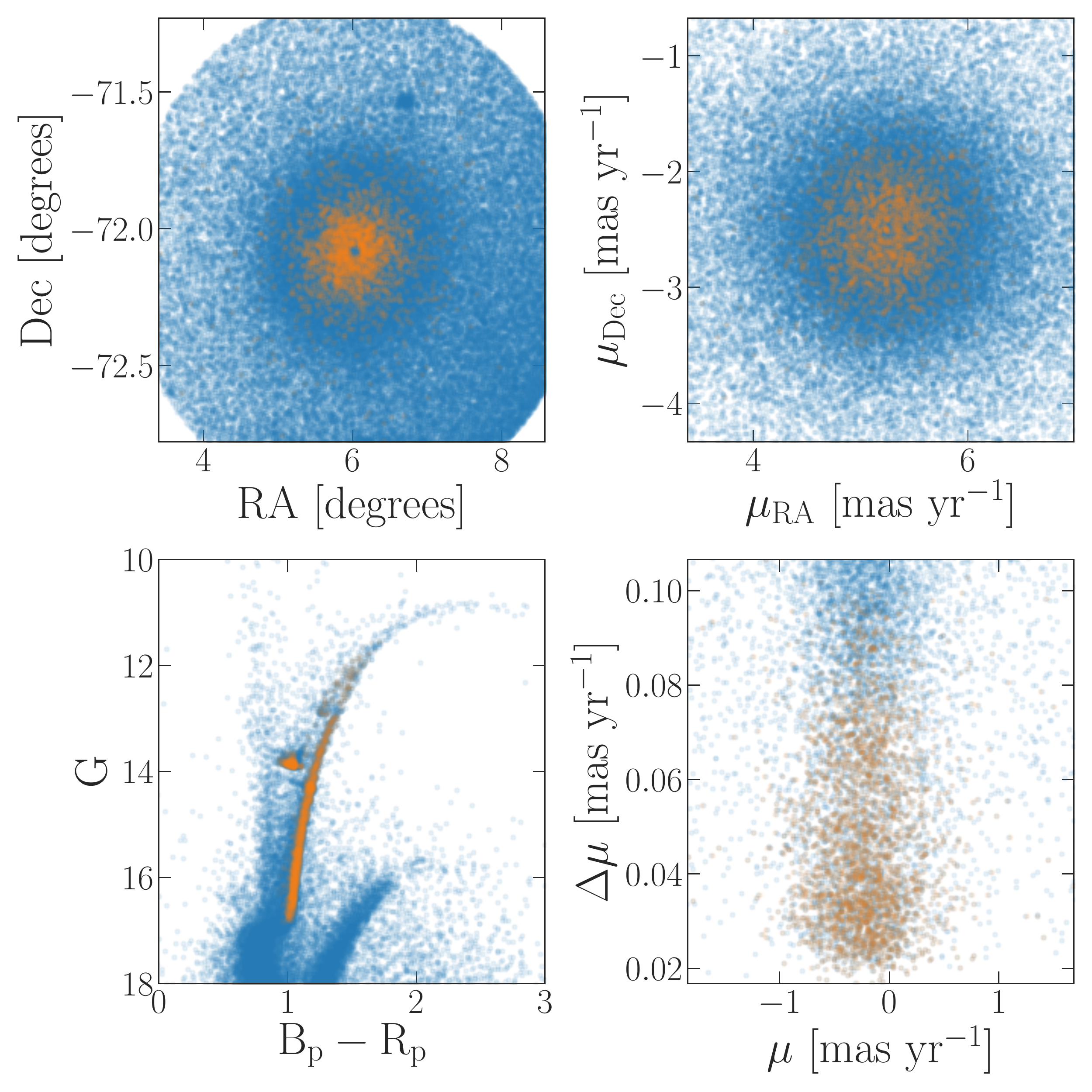}
\caption{Positions (top left), proper motions (top right), colour magnitude diagram (bottom left) and error in total proper motion as a function of total proper motion (bottom right) for \gaia\ sources within the limiting radius of 47 Tuc. Blue points illustrate all stars within the limiting radius of 47 Tuc while orange points show stars that pass our criteria for cluster membership.
\label{fig:cut}}
\end{figure}

\begin{figure*}
\includegraphics[width=\textwidth]{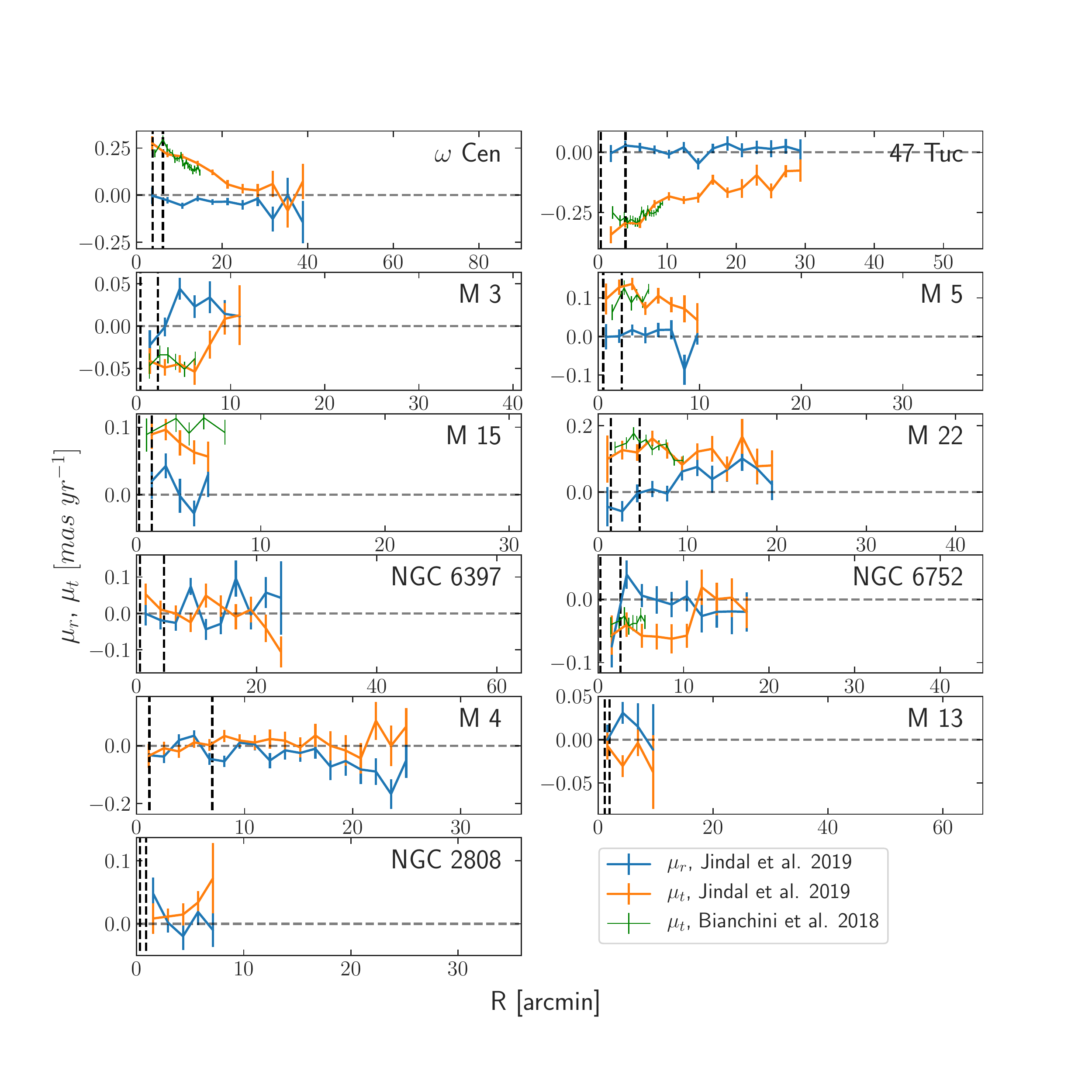}
\caption{Average radial (blue) and tangential (orange) proper motions as a function of radius from the cluster centre, with the vertical dashed lines marking the core (left line) and half-light (right line) radius of the cluster. Average tangential proper motion profiles from \citet{bianchini18} (green) are shown for comparison when possible. Dynamically stable clusters should have no net radial motion, except for that due to perspective expansion. The GCs $\omega$ Cen and M 3 have radial motions consistent with the expected perspective expansion due to their significant line-of-sight motion. Tangential proper motion is observed in several radial bins to 3 $\sigma$ confidence for every cluster except M 4, M 13, NGC 2808, and NGC 6397.
\label{fig:pmrt}}
\end{figure*}

We make use of \gaia\ data to study the internal kinematics of 11 GCs. Our criteria for which GCs to study are based on a combination of distance from the Sun and the expected number of member stars. We further avoid clusters that are aligned with the Galactic Bulge where contamination is extremely high. The final list consists of NGC 104 (47 Tuc), NGC 2808, NGC 5139 ($\omega\ {Cen}$), NGC 5272 (M 3), NGC 5904 (M 5), NGC 6121 (M 4), NGC 6205 (M 13), NGC 6397, NGC 6656 (M 22), NGC 6752, and NGC 7078 (M 15). We query the \gaia\ DR2 database for the photometric, spectroscopic, and astrometric quantities to a distance encompassing $r_L$. For the purposes of this study, the centres of each cluster are taken from SIMBAD and their structural parameters are taken from \citet{deBoer19}.

Away from the dense cluster centre, background stars make up a significant fraction of sources in the sky and need to be removed from consideration. Furthermore, objects with unreliable astrometry or photometry are sources of noise that also must be removed. Prescriptions for selection vary widely among recent studies \citep[e.g.][]{milone18, bianchini18, baumgardt18b, vasiliev18}. We attempt to combine a variety of appropriate filters which we list below. For illustrative purposes, the initial raw stellar dataset from the \gaia\ archive is compared with our final stellar dataset for 47 Tuc in Figure \ref{fig:cut}.

\begin{enumerate}[i]

\item \label{item:radeccut}
As previously mentioned, \citet{deBoer19} used data from \gaia\ to determine the density profiles of 81 GCs, which include the 11 considered in this study. The authors also fit lowered isothermal models \citep{gieles15} to the profile to estimate the limiting radius $r_L$ of each cluster. We use $r_L$ as an initial spatial cut in RA and Dec (see the top left panel of Figure \ref{fig:cut}).

\item \label{item:pmcut}
In most cases, GCs have distinct proper motions from contaminating sources, so proper motions can be used to identify cluster members. We fit a 2D Gaussian to the proper motion distribution and extract the standard deviation and central proper motion. We then only include sources whose proper motions lie within 4 standard deviations of the proper motion centre. The top right panel of Figure \ref{fig:cut} illustrates how for 47 Tuc this proper motion cut removes a significant fraction of stars in the outer regions of the cluster. 47 Tuc is somewhat of a special case, because the Small Magellanic Cloud (SMC) is very close in position space---its outskirts can be seen in the lower right of the top left panel of Figure \ref{fig:cut}---but has a very different proper motion (not visible in the range plotted in the top right panel) that allows SMC stars to be efficiently cut from the sample.

It should be noted that the choice of 4 standard deviations was chosen based on a visual inspection of the proper motion phase space. Cutting objects with significant deviation from the mean proper motion will cause any cluster members that are in the process of escaping the cluster due to evaporation or ejection to be lost. While this may introduce some bias in our kinematic quantities, the effects of contamination become increasingly pronounced with less restrictive cuts.

\item \label{item:cmdcut}
The sources in a GC are generally part of a population of stars with the same age and metallicity, and they therefore trace a particular stellar isochrone tightly on a CMD. We can thus use the sources' distances from this isochrone as a selection tool. We begin by restricting the region of consideration to approximately 3 $r_h$ from the cluster centre such that the isochrone reveals itself sharply. We then trace the boundaries on the left and right side of the isochrone by hand, including a rectangular region around the horizontal branch. All sources that fall outside of these defined boundaries are removed from consideration. In addition, we remove sources that are fainter than $G = 18$, because their astrometry is typically quite poor (see next section). The bottom left panel of Figure \ref{fig:cut} illustrates the CMD of all queried stars and our selected stars in 47 Tuc.

We note that there may be inconsistencies in the CMD cutting between clusters since isochrone boundaries are set manually. However, given that the CMDs reveal fairly sharp boundaries for most clusters and the boundaries are sparsely populated, any error associated with our CMD cutting will be minimal. In particular, this should not lead to any kinematic bias in our sample, as position on the CMD is not correlated with kinematics. Hence we expect that the results should not change significantly based on our CMD cuts alone. 

\item \label{item:pmerrorcut}
We remove data with poor astrometry from the datasets. We analytically propagate the uncertainties in both components of the proper motions to get the uncertainty in the overall proper motion magnitude ($\Delta \mu$), and set a maximum threshold such that the observational error was always less
than the dispersion in the total proper motion of the stellar population. With the exception of $\omega$ Cen, M 4, and M 22, the total proper motion error was restricted to $\Delta \mu < 0.1\ \mathrm{mas\,yr}^{-1}$. For $\omega\ {Cen}$, M 4, and M 22, we use $\Delta \mu < 0.2\ \mathrm{mas\,yr}^{-1}$ instead due to the higher observational errors present in the inner regions of these clusters. Finally, only sources with a high quality astrometric solution ($\texttt{astrometric\_excess\_noise}<1\ \mathrm{mas}$) were kept \citep[for the definition of this DR2 variable, see][]{lindegren18}. The lower right panel of Figure \ref{fig:cut} shows the total proper motions $\mu$ and the $\Delta \mu$ of background sources and accepted stars in 47 Tuc.

\item \label{item:bprpexcesscut}
Near the centre of a typical GC, sources are very crowded, and multiple sources (especially faint ones) may be considered as one object. Crowding introduces scatter in the photometric solution of these objects, affecting the magnitudes quoted for the blue ($G_\mathrm{BP}$) and red ($G_\mathrm{RP}$) bands, and is characterized by the parameter $E=\texttt{phot\_bp\_rp\_excess\_factor}$. Following the suggestions in \citet{lindegren18}, we add the criterion $E<1.3 + 0.06(G_\mathrm{BP} - G_\mathrm{RP})^2$.
\end{enumerate}

We do not use parallax as a contamination filter, because we find that the CMD cuts already exclude most of the sources that a distance cut would eliminate, and errors in the parallaxes are generally very large in DR2 for sources at the distance of the GCs that we consider. Furthermore, we do not use a line-of-sight velocity filter because DR2 lacks the robust spectroscopy needed to obtain values for most sources. The next \gaia\ data releases will include line-of-sight velocities for additional sources, which will provide another method for constraining membership in heavily crowded and contaminated clusters without relying on a cross-correlation with other observations, which may be missing certain sources.

\section{Results}
\label{sec:resultsdiscussion}


\begin{figure*}
\includegraphics[width=\textwidth]{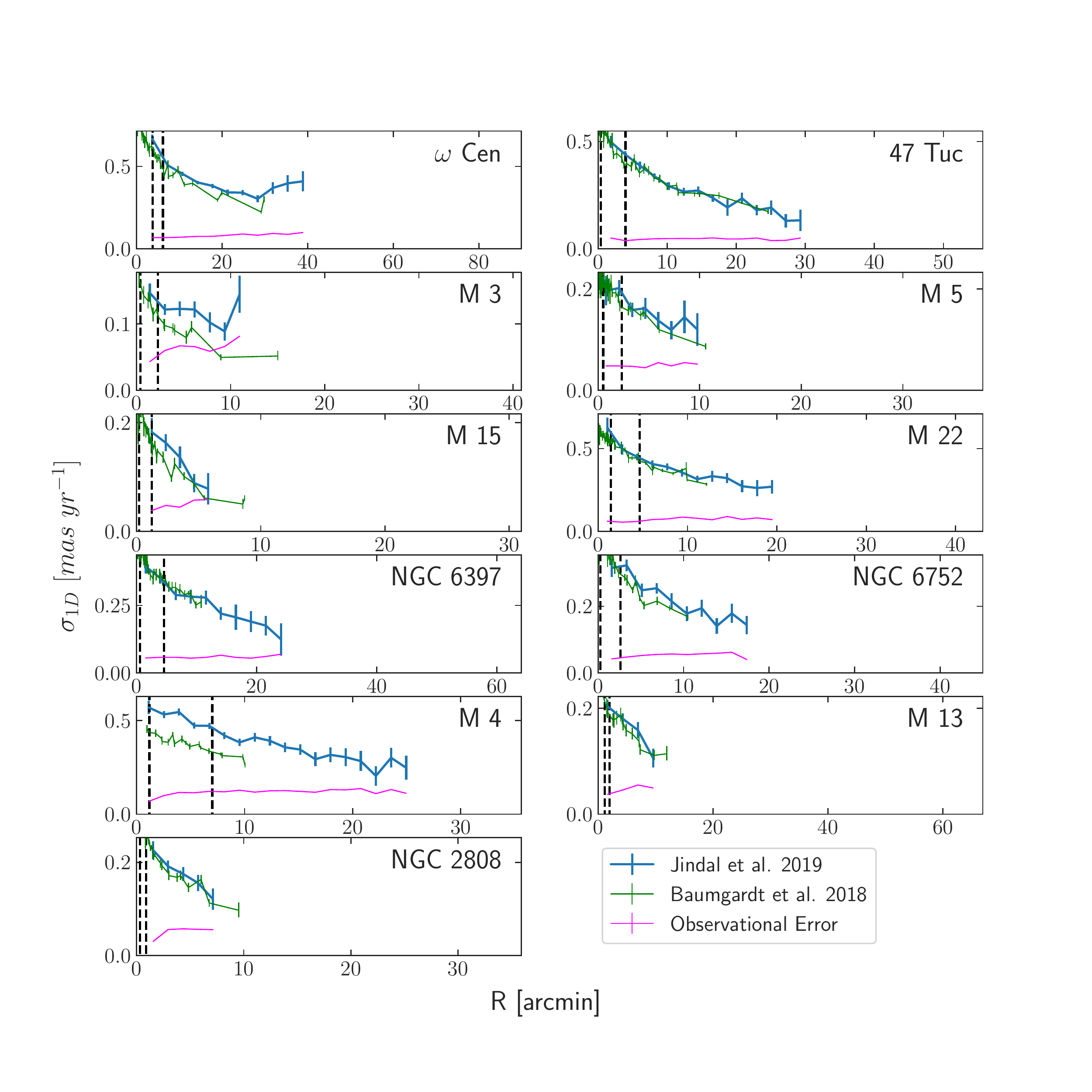}
\caption{Average one-dimensional velocity dispersion as a function of radius from the cluster centre. For comparison, we show the measurements from \citet{baumgardt18b} in green and the contribution from random, observational errors in magenta. The velocity dispersion declines significantly from the centre to the outskirts in all clusters. 
\label{fig:disptotal}}
\end{figure*}

\begin{figure*}
\includegraphics[width=\textwidth]{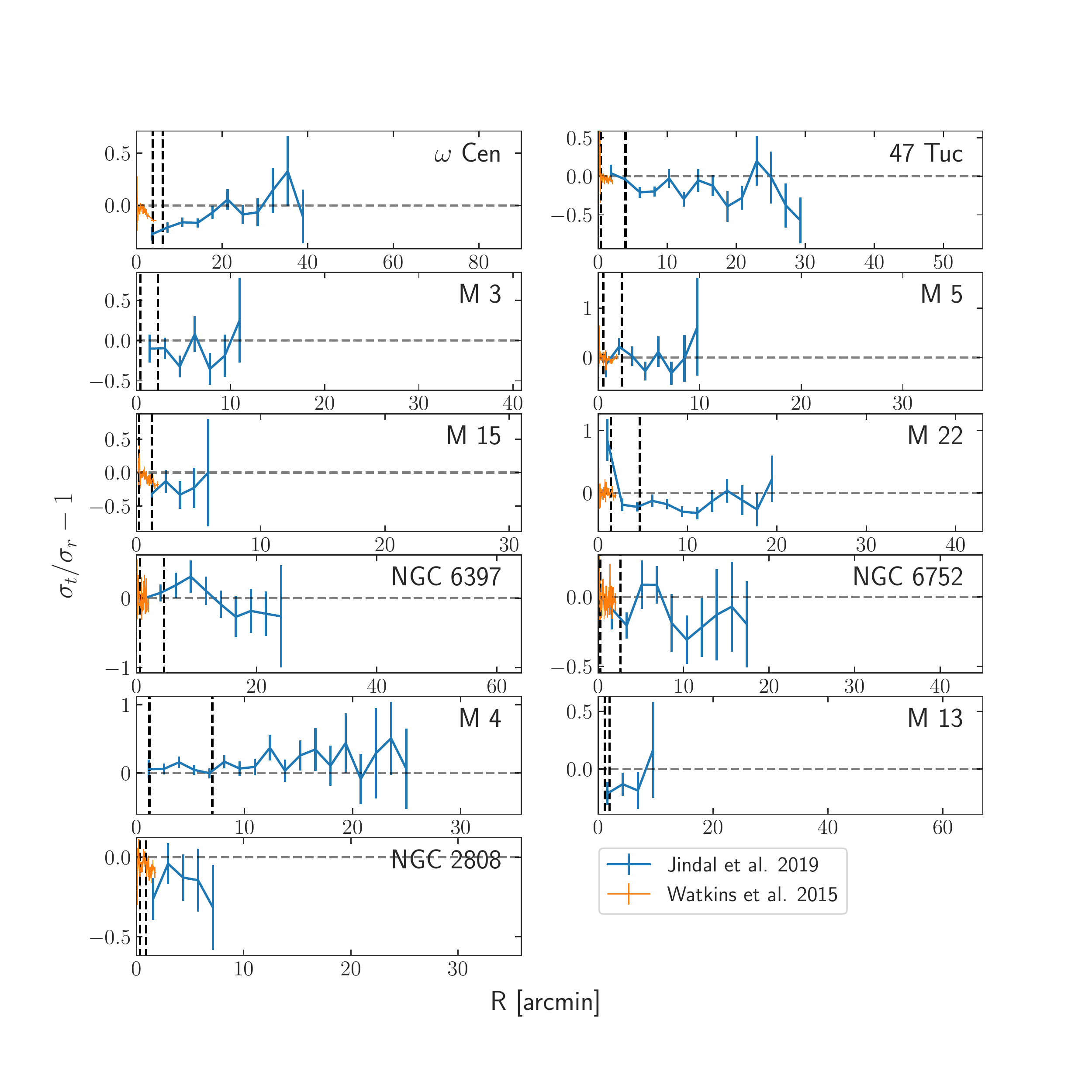}
\caption{Proper motion anisotropy as a function of radius from the cluster centre. $\omega$ Cen, 47 Tuc, M 3, M 13, M 15, and M 22 all show a clear radial anisotropy in their inner parts. NGC 2808 and NGC 6752 both show evidence for weak radial anisotropy, but isotropy can not be ruled out within uncertainty. Overall, the profiles are close to isotropy in the outskirts of all clusters.
\label{fig:anisotropy}}
\end{figure*}

Using the members selected for each GC as described in the previous section, we determine kinematic profiles for each cluster with the available 5-parameter astrometry. We perform an orthographic projection of the usual celestial coordinates and proper motions \citep[see][equation 2]{gaiahelmi18}, which transforms right ascension and declination into $x$ and $y$ coordinates relative to the centre of the clusters. The proper motion of the cluster as a whole is then subtracted. We determine radial profiles of various quantities by binning the entire dataset out to the cluster's limiting radius $r_L$ over 25 bins and computing central bin values as the mean across 500 bootstrap resampling iterations with uncertainties obtained as the standard deviation across these iterations. With the exception of $\omega$ Cen, all bins with more than 10 stars are presented in the following sections. For $\omega$ Cen the radial bins are well populated out to its $r_L$ ($\approx90$ arcmin), but an inspection of the proper motion distribution shows that contamination clearly starts to become significant by 40 arcmin, even with all the cuts described in the previous section. Therefore we have elected to cut the $\omega$ Cen dataset at 40 arcmin.

\subsection{Proper motion profiles}

We first project the proper motions into radial and tangential components $\mu_{r}$, $\mu_{t}$ about the cluster centres and display these profiles in Figure \ref{fig:pmrt}. The radial proper motions are expected to be zero at all radii for dynamically-stable GCs, although they may be non-zero when perspective expansion---the apparent angular contraction or expansion of a cluster due to it moving away from or towards us---due to significant line-of-sight motion is important. The radial proper motion profiles in Figure \ref{fig:pmrt} are all consistent with zero except for M 3 and $\omega$ Cen. Using the average line-of-sight velocities presented in \citet{Baumgardt18a}, the expected radial proper motion due to perspective expansion for these two clusters \citep[illustrated by][]{vasiliev18} is consistent with our results.

The tangential proper motion profiles in Figure \ref{fig:pmrt} are also expected to be consistent with zero, unless the cluster rotates. We detect tangential motion with confidence greater than 3$\sigma$ in several radial bins for every cluster except M 4, M 13, NGC 2808, NGC 6397. 47 Tuc and $\omega$ Cen in particular show significant rotation, the former being in agreement with \citet{milone18}. In each case, rotation weakens steadily with increasing radius and approaches zero well before the GC's $r_L$ (although uncertainties generally also become very large at these radii). Profiles taken from \citet{bianchini18} are also overplotted for comparison; these are consistent with our results. Because \citet{bianchini18} focused on the inner regions of clusters only, our profiles go out to significantly farther radii. Rotation is clearly detected at these extended radii for almost half of the clusters, with no rotation found in the outer regions M 3, M 4, M 13, NGC 2808 and NGC 6397.

We note that in the case where the rotation axis lies in the plane of the sky, proper motions will not reveal rotation in the radial profiles due to the geometry of the situation. Thus, where no rotation is detected by our methods (e.g. NGC 6397), rotation cannot be ruled out. For a proper analysis at the individual cluster level, line-of-sight velocities are required to supplement proper motions and provide 3D kinematics.

\subsection{Proper motion dispersion profiles}

As a measure of the proper motion dispersion in each cluster, we calculate the one-dimensional proper motion dispersion ($\sigma_{1D}$) as a function of radius, where $\sigma_{1D}$ is defined as the mean of the radial and tangential velocity dispersion components. To not be adversely affected by a few outlying proper motions, we first compute a robust estimate of the dispersion as the median absolute deviation of the radial and tangential proper motion in each radius bin, with central values and uncertainties therein given by the means and standard deviations (respectively) across the same 500 bootstrap iterations described above; we multiply the median absolute deviation with 1.4826 such that for a Gaussian distribution it would equal the standard deviation. Finally, it should be noted that in the $\sigma_{1D}$ profiles illustrated in Figure \ref{fig:disptotal}, uncertainties have been propagated analytically and we have subtracted in quadrature the contribution of the observational error.

Our results show a steadily decreasing dispersion with radius, as expected due to the decreasing density and increasing tidal field strength in the outer regions. The random, observational uncertainty in measured proper motion components is analytically propagated to an uncertainty in the total proper motion, and the median of this value in each bin is shown in magenta. This uncertainty generates a source of velocity dispersion in addition to the intrinsic dispersion. While we subtract this observational uncertainty in quadrature to estimate the intrinsic dispersion, it is important to note that the intrinsic dispersion is in most bins well above the random, observational uncertainty in all of the clusters. Therefore, the observational uncertainty does not adversely influence our results.

The measured one-dimensional velocity dispersion profiles (Figure \ref{fig:disptotal}) all agree with those presented in \citet{baumgardt18b}, except for a slight tension for M 3 and M 4. For these two clusters, we find that we can restrict our cluster members to those with brighter magnitudes or smaller errors in total proper motion (see items \ref{item:cmdcut}, \ref{item:pmerrorcut} in Sec.~\ref{sec:selection}) until our results agree with \citet{baumgardt18b}. However, this approach would cause a significant fraction of cluster members to be lost. Hence our dataset includes stars that \citet{baumgardt18b} discarded as cluster members based on their selection criteria. It is important to point out that calculating $\sigma_{1D}$ with member stars from \citet{deBoer19} with high membership probabilities ($>90\%$) are in disagreement with both \citet{baumgardt18b} and this work. All three works predict similar inner region valeus of $\sigma_{1D}$ in M 3, but stars from \citet{deBoer19} predict a much steeper decrease in $\sigma_{1D}$ with radius than \citet{baumgardt18b}  or this work. Conversely in M 4, \citet{deBoer19} stars yield a higher inner $\sigma_{1D}$ before converging to the $\sigma_{1D}$ profile we measure beyond 5 arcminutes. These differences highlight the difficulty involved in confirming cluster members with Gaia data and the importance of the selection criteria details used by different studies.


In all cases, we find that the steady decrease in dispersion found by \citet{baumgardt18b} continues out to extended radii, except in $\omega$ Cen, where there a previously-reported upturn begins. Similar trends are seen using member stars taken from \citet{deBoer19} with membership probabilities greater than $90\%$ in $\omega$ Cen, 47 Tuc, and M 15. Given the low density of stars in the outer regions of these clusters, and the fact that the upturn occurs in both the radial and tangential velocities dispersion, the upturn is likely due to a combination of having a small sample size and contamination. If the upturn was due to an interaction with the tidal field, like a tidal shock, the upturn would only be visible in one of the radial (if the shock was on-going or very recent) or tangential (after stars energized to radial orbits have escaped the cluster) velocity dispersion profiles, not both.  

\subsection{Anisotropy profiles}

Finally, we compute the anisotropy parameter $\beta = \sigma_{t}/\sigma_{r} - 1$, illustrated in Figure \ref{fig:anisotropy}. A positive value indicates an excess in the tangential velocity dispersion, while a negative value indicates an excess in the radial velocity dispersion. Uncertainties are again propagated analytically. For comparison purposes, we overplot the profiles measured by \citet{watkins15} using HST data on the inner regions when possible. While limited to the inner $\approx 100$ arcsec of the cluster, data from HST is significantly more accurate due to its higher resolution and lower magnitude limit.

The only profiles that appear to be in disagreement with \citet{watkins15} are those of $\omega$ Cen and M 22, and only for the innermost radial bins. These disagreements can easily be attributed to \gaia\ having difficulty resolving the dense cores of GCs; most inner region stars have high proper motion uncertainties and poor photometric solutions, and we therefore remove them from our dataset. There is no disagreement if stars from \citet{deBoer19} with $90\%$ membership probability are included in the datasets.

Almost all of the clusters show signs of being isotropic in their cores, radially anisotropic in their inner regions, and again isotropic in their outskirts. $\omega$ Cen, 47 Tuc, M 3, M 13, M 15 and M 22 in particular all show evidence for at least a small amount of radial anisotropy in their inner regions. The large uncertainties in measurements of $\sigma_r$ and $\sigma_t$ in NGC 2808 and NGC 6752 make it difficult to determine whether their inner regions are consistent with weak radial anisotropy or if they are isotropic. We find that M 4, M 5, and NGC 6397 are primarily isotropic.

For $\omega$ Cen, anisotropy has been previously investigated using HST proper motions by \citet{vandermarel10}. They report a significant detection of a steadily increasing radial anisotropy up to 5 arcmin, which is consistent with our results (although our error bars are large in these inner regions). We measure a return to isotropy at about 20 arcmin, which then holds out to at least 40 arcmin. Our finding for 47 Tuc is consistent with the results of \citet{milone18}, who present data up to 18 arcmin. Beyond this distance, 47 Tuc is primarily isotropic out to the limit of our dataset (nearly 40 arcmin or 2/3 $r_L$). 

The profiles of M 3, M 13 and M 15 are both extended significantly compared to previous works, with some tension around where the transition from an isotropic core to radial anisotropy occurs in M 3 and M 13. \citet{gunn79} was the first to suggest that the core of M 3 was isotropic with a transition to anisotropy at 5 core radii $r_c$, which was later confirmed by \citet{cudworth79} (see also \citealt{kamann14}). In slight disagreement with these previous works, we find that M 3 does not return to isotropy until at least 10 arcminutes (20 $r_c$). Similarly, early work by \citet{lupton87} found that M 13 was isotropic out to 5 $r_c$ before becoming anisotropic. While we do not resolve the core of M 13 here, we find evidence for anistropy in M 13 as early as 1 $r_c$. In both cases the discrepancy can be attributed the significantly improved quality of the \gaia\ data. Radial anisotropy was also previously reported for M 15 by \citet{watkins15}, out to the limit of their dataset ($\approx 1/ r_h$). We find that the anisotropy continues until approximately $4 r_h$, where the cluster returns to isotropy.

Finally, radial anisotropy is detected for the first time in M 22. Similar to M 3, \citet{watkins15} found that the core of M 22 is isotropic. Continuing on outside of the core with \gaia, we find that within 2 core radii the cluster develops radial anisotropy. It is not until 10.5 core radii (or 3 $r_h$) does the cluster isotropize.

\section{Discussion and Conclusions} \label{sec:conclusion}

We have shown that astrometry from \gaia\ DR2 reveals interesting internal dynamics in the plane of the sky for 11 nearby GCs. To make the measurements presented in this paper, we applied a detailed, four-step procedure to select appropriate sources for analysis. We  remove the extensive contamination from background sources in the intermediate and outer parts of the studied GCs by selecting members based on their location in proper motion space and on the CMD. We also filter out sources with unreliable data by constraining the quality of the astrometry and photometry. These steps are outlined in items \ref{item:pmcut} to \ref{item:bprpexcesscut} in Sec.~\ref{sec:selection}. 

We list the main results and points for discussion below. 
\begin{itemize}

\item
We detect rotation at high significance for 7 clusters at various radii (Figure \ref{fig:pmrt}). We find a steady decrease in tangential proper motion $\mu_{t}$ with radius, demonstrating that GCs rotate differentially. Our profiles agree very well with the literature in the range where they overlap, but we probe rotation at radii that are significantly larger than previously considered. This allows us to detect rotation in the outer regions of $\omega$ Cen, 47 Tuc, M 5, M 22, NGC 6752 for the first time.

\item
The one-dimensional proper motion dispersion $\sigma_{1D}$ also shows a steady decrease with radius (Figure \ref{fig:disptotal}). These profiles are generally consistent with literature measurements, and the slight tension that exists for M 3 and M 4 can be explained by the observational uncertainties and different selection criteria. We probe radii that are significantly larger than previous studies for $\omega$ Cen, M 4, M 22, NGC 6397 and NGC 6752.

\item

We determine the proper motion anisotropy profiles of each cluster (Figure \ref{fig:anisotropy}). We detect excess velocity dispersion in the radial direction in the inner regions of $\omega$ Cen, 47 Tuc, M 3, M 13, and M 15, each of which are mostly consistent with the literature, but we extend these detections to larger radii. For M22, we present the first detection of radial anisotropy outside of its isotropic core. In each case, the observed anisotropy eventually disappears and the outer regions of these clusters are isotropic. Anisotropy cannot be confirmed or ruled out in NGC 2808 or NGC 6752, due to the uncertainties involves, while M4, M5, and NGC6397 are all consistent with being isotropic.

\end{itemize}

\begin{figure}
\includegraphics[width=\columnwidth]
{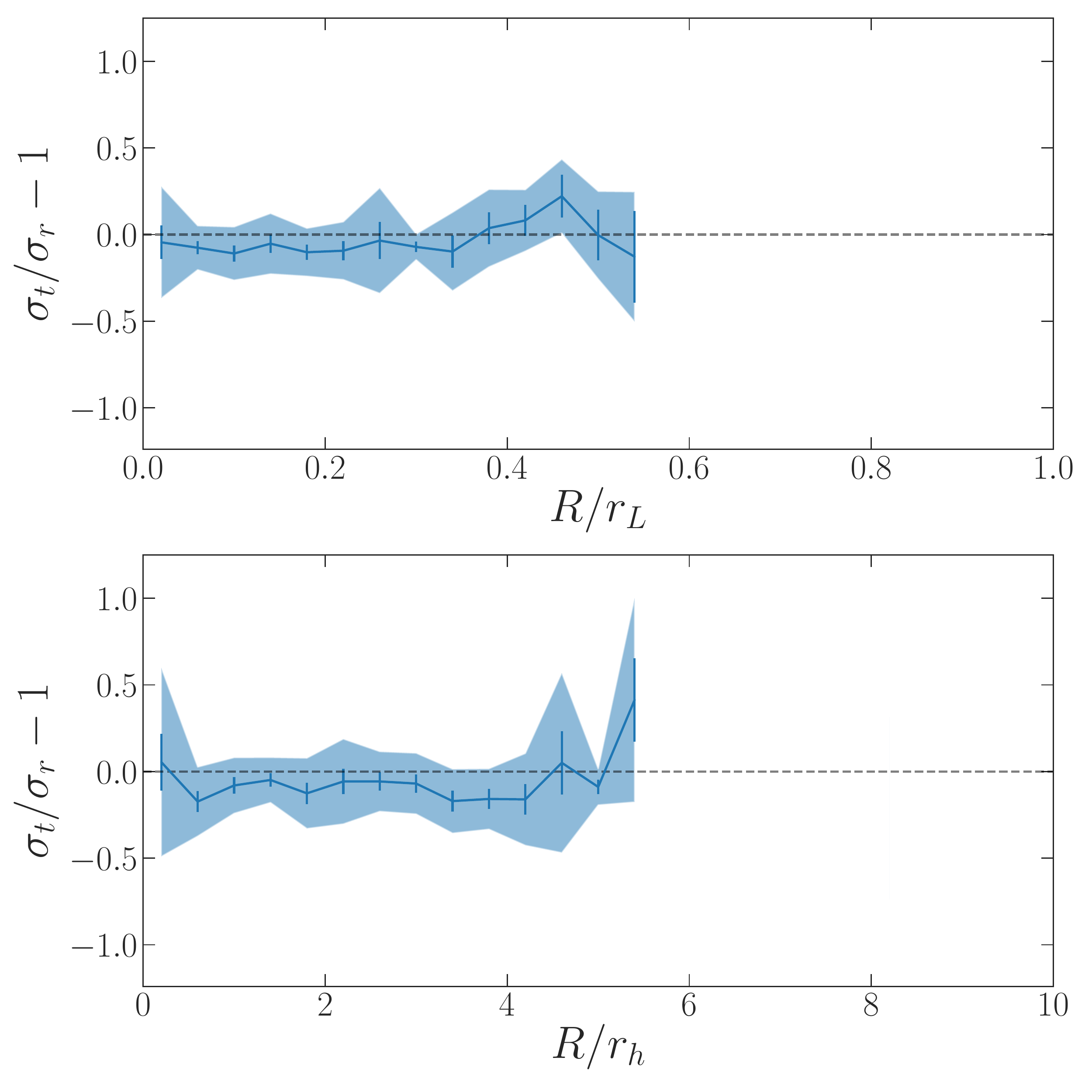}
\caption{The combined mean anisotropy profile of all 11 clusters when normalized by the limiting radius (top panel) and the half-light radius (bottom panel). Error bars represent the error in the mean profile while the shaded region shows the scatter (standard deviation) of different clusters' profiles about the mean profile.}

\label{fig:med_anisotropy}
\end{figure}

Combining the anisotropy profiles of all 11 clusters into one mean profile in Figure \ref{fig:med_anisotropy}, scaled by either $r_h$ or $r_L$, clearly illustrates that outside of their core, clusters are primarily radially anisotropic out to at least $4\ r_h$ or $0.35\ r_L$. Beyond these radii, clusters are isotropic out to the limits of our datasets. Such an anisotropy profile is consistent with clusters forming tidally under-filling, such that radial anisotropy develops as the cluster expands. The core of the cluster is able to quickly isotropize due to its short relaxation time \citep{lutzgendorf11}, while a return to isotropy in the outer regions of these clusters is a direct result of external tides stripping stars on eccentric orbits \citep{zocchi16,tiongco16}. Future \gaia\ data releases will allow for the dataset to be extended beyond the bright and nearby GCs considered in this investigation in order to determine whether or not this average anisotropy profile is applicable to all clusters.

The cases where radial anisotropy is observed in the inner regions of the cluster ($\omega$ Cen, M 3, M 13, M 15 and 47 Tuc) are likely examples of clusters that either formed tidally under-filling, are dynamically young, have undergone core collapse, or are strongly affected by the presence of multiple populations \citep{tiongco16,zocchi17,milone18}. $\omega$ Cen has one of the longest relaxation times of any cluster \citep{harris96}. Hence, it is dynamically young and relaxation is not expected to have erased the anisotropy profile that young clusters are expected to develop. Only in the outer regions of the cluster, where the tidal field has been able to isotropize the cluster, has anisotropy been erased. It is interesting to note, however, that anisotropic cluster models presented by \citet{zocchi17} find that the degree of radial anisotropy in $\omega$ Cen should increase up to roughly 30-40 arcmin, after which the system should make a sharp turn toward isotropy due to tidal torque introducing isotropy in the velocity dispersion \citep{oh92} and mass loss due to dynamical interactions \citep{giersz97}. Our finding that the cluster returns to isotropy by 20 arcmin, likely due to tidal interactions, suggests that either $\omega$ Cen's anisotropy radius is much smaller than considered by \citet{zocchi17} or the cluster is not well fit by the lowered isothermal models presented in \citet{gieles15}. Similar to $\omega$ Cen, M 3 has a very long relaxation time (6.2 Gyr) and is therefore expected to be dynamically young. It is, however, more tidally filling than $\omega$ Cen so the earliar return to isotoropy is not surprising.

M 13 and M 15 are both clusters that are tidally under-filling at their current location, but tidally filling at pericenter. Hence these clusters were even more under-filling in the past and were therefore able to develop a strong anisotropy profile early in their dynamical lifetimes \citep{tiongco16}. While tides have likely helped remove some stars on radial orbits in the outer regions of the clusters, the majority of the cluster has remained tidally unaffected. It is interesting to note that both clusters have intermediate relaxation times ($\approx 2$ Gyr). Therefore only weak radial anisotropy is expected in these clusters, as observed, since relaxation is able to somewhat decrease the degree of anisotropy in the inner regions over time.

Finally, the anisotropy observed in 47 Tuc has been attributed to the outward migrated of the so-called second population stars (sometimes referred to as second generation), which are believed to form more centrally concentrated than the first population stars \citep{milone18}. A kinematic study of multiple populations in $\omega$ Cen and M 15 may also reveal that the radial anisotropy is only due to the second generation of stars, though additional investigation would be needed to make such conclusions. Furthermore, given the fact that the properties of $\omega$ Cen are notoriously complex, which has led to suggestions that it is not actually a GC but the surviving core of a dwarf galaxy \citep{gnedin02}, explaining the anisotropy profile of $\omega$ Cen may not be straightforward.

For the three clusters where no radial anisotropy is observed, it is difficult to determine whether the lack of anisotropy is due to the clusters forming tidally filling (such that radial anisotropy never developed) or due to the combined effects of relaxation and tidal stripping. However, the current orbits \citep{gaiahelmi18} and half-mass relaxation times ($t_{rh}$) \citep{harris96} of these remaining clusters provide some insight. For example, M 4 and NGC 6397 are tidally filling and have $t_{rh} < 1$ Gyr, which suggests both tides and relaxation have shaped their anisotropy profiles. In the case of M 4 specifically, tides are more likely responsible for shaping the cluster's anisotropy profile than relaxation as its pericentre is approximately 0.4 kpc \citep{gaiahelmi18}. Finally, M 5 is tidally filling with an intermediate relaxation times $t_{rh} > 1.6$ Gyr. Hence the cluster either formed tidally filling and never developed any radial anisotropy in the first place or strong tides have helped remove stars on primarily radial orbits.

For the two clusters where anisotropy cannot be confirmed or ruled out, we again make use of their orbits and structural properties to guide our expectations. NGC 6752 is tidally under-filling (at its current position and at pericentre) and has a short $t_{rh}$ ($\approx 800$ Myr), indicating that a lack of anisotropy would not be surprising as the cluster is strongly affected by relaxation. A similar argument can be made for NGC 2808, although its half-mass relaxation time is slightly larger and it is tidally filling at pericentre. Furthermore, NGC 2808 may also be strongly influenced by the presence of multiple stellar populations \citep[e.g.][]{carretta18}, which may lead to some degree of internal radial anisotropy despite the cluster being dynamically old and tidally affected.

It is clear that the vast quantity of accurate astrometry available from \gaia\ allows for unprecedented insights into the kinematic behaviours of globular clusters. As relics of dynamical interactions from the early Milky Way, understanding these objects will play a crucial role in understanding the behaviour of the Galaxy as a whole. Future data releases from \gaia\ will help further our ability to constrain the radial profiles of key kinematic properties in a larger sample of GCs covering a wider range of GC orbits, tidal-filling factors, and internal relaxation times.

\section*{Acknowledgements}

This study received support from the Natural Sciences and Engineering Research Council of Canada (NSERC; funding reference number RGPIN-2015-05235) and from an Ontario Early Researcher Award (ER16-12-061). JW also acknowledges funding through an NSERC Postdoctoral Fellowship. This work has made use of data from the European Space Agency (ESA) mission \gaia (\url{https://www.cosmos.esa.int/gaia}), processed by the \gaia Data Processing and Analysis Consortium (DPAC, \url{https://www.cosmos.esa.int/web/gaia/dpac/consortium}). Funding for the DPAC has been provided by national institutions, in particular the institutions participating in the \gaia Multilateral Agreement.






\bibliographystyle{mnras}
\bibliography{ref2}



\appendix
\renewcommand\thefigure{\thesection{A}\arabic{figure}}

\setcounter{figure}{0}


\bsp	
\label{lastpage}
\end{document}